# THE CONFLUENT SYSTEM FORMALISM:
# I. THE MASS FUNCTION OF OBJECTS
# IN THE PEAK MODEL

Alberto Manrique

Departament d'Astronomia i Meteorologia, Universitat de Barcelona,
Avda. Diagonal 647, E-08028 Barcelona, Spain
e-mail: alberto@faess2.am.ub.es

and

Eduard Salvador-Solé

Departament d'Astronomia i Meteorologia, Universitat de Barcelona,
Avda. Diagonal 647, E-08028 Barcelona, Spain
e-mail:eduard@faess0.am.ub.es

## ABSTRACT

This is the first paper of a series of two devoted to develop a practical method to describe the growth history of bound virialized objects in the gravitational instability scenario without resorting to $N$-body simulations. Here we present the basic tool of this method, "the confluent system formalism", which allows us to follow the filtering evolution of peaks in a random Gaussian field of density fluctuations. This is applied to derive the theoretical mass function of objects within the peak model framework. Along the process followed for the derivation of this function, we prove that the Gaussian window is the only one consistent with the peak model ansatz. We also give a well justified derivation of the density of peaks with density contrast upcrossing a given threshold in infinitesimal ranges of scale and correct this scale function for the cloud-in-cloud effect. Finally, we characterize the form of the mass vs. scale and the critical overdensity vs. collapse time relations which are physically consistent with the peak model in an Einstein-de Sitter universe with density field endowed with different power spectra. The result is a fully justified semianalytical mass function which is close to the Press & Schechter (1974) one giving good fits to $N$-body simulations. But the interest of the confluent system formalism is not merely formal. It allows us to distinguish between accretion and merger events, which is essential for the detailed modelling of the clustering process experienced by objects.

*Subject headings:* cosmology: theory – galaxies: clustering – galaxies: formation

## 1. INTRODUCTION

The theoretical mass function of bound virialized objects plays a crucial role in many cosmological issues. For the last twenty years, a considerable effort has been devoted to properly infer it in the gravitational instability scenario of galaxy formation from the statistics of the primordial random density field. The most widely used expression for such a mass function is due to Press & Schechter (1974, PS):

$$N(M,t)\,dM \;=\; 2\,\frac{\rho}{M}\,\left|\frac{\partial V(\delta_c, R)}{\partial R}\right|\,\frac{dR}{dM}\,dM. \qquad (1)$$

In equation (1), $\rho$ is the mean mass density of the universe at the arbitrary initial epoch $t_i$ after recombination when fluctuations are still linear, with growing factor only dependent on time, and Gaussian distributed. (Note that, in comoving lengths as assumed throughout the present paper, $\rho$ takes its current value.) The function

$$V(\delta_c, R) \;=\; \frac{1}{2}\,\mathrm{erfc}\!\left[\frac{\delta_c}{\sqrt{2}\sigma_0(R)}\right] \qquad (2)$$

gives the volume fraction of points with density contrast $\delta$, defined as the density fluctuation normalized to the mean density, exceeding some positive linear threshold $\delta_c$ in the density field smoothed by means of a top hat window of scale $R$, $\sigma_0(R)$ is the rms $\delta$ on this scale. The function $M(R) = 4\pi/3\,\rho\,R^3$ relates the mass of objects with the filtering scale, and the dependence on $t$ in the right-hand member of equation (1) is given, in an Einstein-de Sitter universe, by $\delta_c(t) = \delta_{c0}\,a(t_i)/a(t)$, with $a(t)$ the cosmic expansion factor and $\delta_{c0}$ a well-known constant approximately equal to 1.69.

Expression (1) is based on the linear extrapolation, for the growing mode, of the growth of density fluctuations in the initial Gaussian random field with limit given by the spherical collapse model. According to this latter model, the collapse time for a shell of radius $R$ around the center, located at $\mathbf{r}$, of a spherically symmetric, outwards decreasing (to avoid shell crossing), linear density fluctuation at $t_i$ only depends on the mean value of $\delta$ inside it. More exactly, the value of the average density contrast for collapse at $t$ in an Einstein-de Sitter universe is $\delta_c(t) = \delta_{c0}\, a(t_i)/a(t)$. Of course, the collapse of the shell of radius $R$ represents the appearance, at $t$, of a virialized object of mass equal to $4\pi/3\,\rho\,R^3$ to 0th order in $\delta_c$. This therefore suggests that any point in the real density field at $t_i$ smoothed with a top hat filter of scale $R$ with density contrast *above* a positive linear threshold $\delta_c$ should tend to collect matter so to reach, at a time $t$ related to $\delta_c$ through the previous expression, a mass *M larger than* $4\pi/3\,\rho\,R^3$. Consequently, by differentiating the volume occupied by such points over $M$ one should obtain the volume contributing at $t$ with objects of mass between $M$ and $M + dM$, and by dividing the result by $M/\rho$ to the number of such objects. Actually, since $V$ in equation (1) is not a volume but a volume fraction, $N(M,t)\,dM$ is a number density. The factor two in the right-hand member of equation (1) is necessary for this mass function to be correctly normalized, that is, for the integral of $M$ times the mass function (1) to be equal to the mean density of the universe. Indeed, every particle in the universe must be at any time $t$ within some virialized object with the appropriate mass.

But the PS mass function is not fully satisfactory. The origin of the "fudge factor two" is unclear and the disappearance of objects of any given mass swallowed by previously collapsed ones owing to cloud-in-cloud configurations is not accounted for. In addition, the real density field is not spherically symmetric and outwards decreasing around any point. As a consequence, the growth of density fluctuations leaving the linear regime deviates from spherical collapse and involves complicated non-local, nonlinear, dynamics. Therefore, it is by no means obvious that the PS prescription can provide a good description of the formation of bound virialized objects. In particular, small changes in those aspects the most strongly connected with the spherical collapse model might be suitable. This leads to the following questions. What is the filtering window that better reproduces the clustering of objects? What is the mass to be associated with the filtering scale $R$? What overdensity does really correspond to the collapse time $t$? Finally, there is no reason for every point above the threshold overdensity to tend to accrete matter. This is expected to happen rather onto density maxima or "peaks" (Doroshkevich 1970; Kaiser 1984; Doroshkevich & Shandarin 1978; Peacock & Heavens 1985; Bardeen et al. 1986, herein BBKS; Bernardeau 1994). It is true that $N$-body simulations seem to show that there is no good correspondence, either, between peaks and objects (Katz, Quinn, & Gelb 1993). But this result may be due to the uncorrected cloud-in-cloud effect, the possible use of an unappropriated window, and the inclusion of density constrasts and masses which do not actually correspond to the collapse times and filtering scales analyzed. In any event, peaks are the best seeds of virialized objects we can think about right now and certainly better physically motivated than the undefined volumes used in the PS prescription.

Yet, the PS mass function gives very good fits to the "empirical" mass function inferred from $N$-body simulations (Nolthenius & White 1987; Efstathiou et al. 1988; Efstathiou &

Rees 1988; Carlberg & Couchman 1989; White et al. 1993; Bahcall & Cen 1993; Lacey & Cole 1994). For this reason numerous authors have tried to properly justify it by introducing slight modifications if necessary. The origin of the fudge factor two and the cloud-in-cloud problem have been solved by Bond et al. (1991) by means of the powerful "excursion set formalism" (and the use of the k-sharp window); see also Jedamzik (1994) for an alternate solution to these problems. The effects of the departure from spherical collapse have also been studied (Monaco 1994, and references therein). But all these improvements only apply to the PS original prescription dealing with undefined regions above the threshold overdensity, while there is no satisfactory derivation of the theoretical mass function for peaks as seeds of virialized objects.

The peak model ansatz states that objects at a time $t$ emerge from peaks with density contrast equal to a fixed linear overdensity $\delta_c$ in the smoothed, on any scale $R$, density field at the arbitrary initial time $t_i$. The critical overdensity is assumed to be a monotonous decreasing function of $t$, while the mass $M$ of collapsing clouds associated with peaks is assumed to be a monotonous increasing function of $R$. (The collapsing cloud associated with a peak is simply the region surrounding the peak with total mass equal to that of the final virialized object at $t$.) Therefore, the evolution, with shifting density contrast, of the filtering scale of peaks at $t_i$ is believed to trace the growth in time of the mass of objects. Of course, this is just an ansatz whose validity has to be assessed, a posteriori, by comparing the clustering model it yields with $N$-body simulations. Note, in particular, that peaks might be good seeds of virialized objects and the mass of their associated collapsing clouds not be just an increasing function of $R$ (see, e.g., Bond 1989) or the time of collapse of such clouds not be just a decreasing function of the smoothed density contrast. These assumptions intended to constrain the freedom left by the unknown dynamics of the collapse of density fluctuations are, nonetheless, very reasonable. They are suggested by the spherical collapse model, approximation which is particularly well suited when dealing with peaks. On the other hand, they are much less restrictive than the specific relations $\delta_c(t)$ and $M(R)$ predicted by that simple model. So there is much room left for any actual departure from it. Finally, there is the extra freedom arising from the filter used, which can be different from the top hat one.

The direct extension of the PS prescription to the peak model suggests itself. The resulting mass function is (Colafrancesco, Lucchin, & Matarrese 1989; Peacock & Heavens 1990)

$$N(M,t)\,dM = A\frac{1}{M}\left|\frac{\partial[n_{pk}(\delta_c,R)\,M_{pk}(\delta_c,R)]}{\partial R}\right|\frac{dR}{dM}\,dM, \qquad (3)$$

where A is a normalization factor, $n_{pk}(\delta_c, R)$ is the number density of peaks with density contrast above the threshold $\delta_c$ in the density field smoothed on scale $R$, and $M_{pk}(\delta_c, R)$ is the average mass of objects emerging from these peaks (when divided by $\rho$, equal to the mean volume subtended by the corresponding collapsing clouds). Note that since peaks included in $n_{pk}(\delta_c, R)$ do not have, in general, density contrast equal to $\delta_c$, the average mass of their collapsing clouds, $M_{pk}(\delta_c, R)$, will differ from $M(R)$. The above mentioned problem with the normalization of the PS mass function and the cloud-in-cloud effect is reflected in the variety of expressions found in the literature for the factor $A$ and the function $M_{pk}(\delta_c, R)$ in equation (3). A more serious problem, however, is that this equation states that the mass in objects emerging from peaks with density contrast upcrossing $\delta_c$ in the range of scales between $R$ and $R + dR$ is equal to the variation from $R$ to $R + dR$ of the mass associated with peaks with density contrast above $\delta_c$. It is therefore implicitly assumed that 1) the total mass associated with peaks (with $\delta > 0$) is conserved with varying scale, and 2) the density contrast of peaks is a decreasing function of scale. Both points seem to follow, indeed, from the peak model ansatz. But this is actually not true. As shown below, point 2 crucially depends on the shape of the window used, while the frequent discontinuities in peak trajectories in the $\delta$ vs. $R$ diagram yielded by mergers invalidate point 1 and, hence, equation (3) in any event.

Before we proceed further, one brief comment on the notation used throughout the paper is in order. Rather than the integrated density of peaks on scale $R$, $n_{pk}(\delta_c, R)$, appearing in equation (3) we will use the differential density of peaks on a fixed scale $R$ with scaled density contrasts $\nu \equiv \delta/\sigma_0(R)$ in an infinitesimal range, denoted by $\mathcal{N}_{pk}(\nu, R)\,d\nu$. (Note that this is the same notation as in BBKS except for the fact that the value of the filtering scale used is explicitly quoted as one parameter.) In addition, we will introduce the differential density of peaks with fixed density contrast $\delta$ on scales in an infinitesimal range, denoted by $N_{pk}(R, \delta)\,dR$ (with the fixed value of $\delta$ as one parameter.) Caution must be made in not mixing up these two densities, as well as their respective conditional forms.

Thus, the only reliable strategy to derive the mass function of objects in the peak model framework is to directly count the density of peaks with density contrast $\delta_c$ in infinitesimal ranges of scale, $N_{pk}(R, \delta_c)\,dR$, and then transform it to the mass function of objects at the time $t$, $N(M, t)\,dM$, by using the appropriate $M(R)$ and $\delta_c(t)$ relations. Unfortunately, apart from the uncertainty about these latter two relations and the filter to be used, as well as the cloud-in-cloud effect which is always present, this strategy faces a new important drawback: the differential density of peaks is well-defined for a fixed filtering scale, not for a fixed density contrast. In other words, we only know the form of $\mathcal{N}_{pk}(\nu, R)\,d\nu$ while we need that of $N_{pk}(R, \delta)\,dR$. Bond (1989) proposed the following "reasonable, although not rigurously derivable", expression for the scale function of peaks at fixed density contrast,

$$N_{pk}(R, \delta)\,dR = \mathcal{N}_{pk}(\nu, R)\,\frac{\partial \nu}{\partial R}\,dR, \qquad (4)$$

with $\mathcal{N}_{pk}(\nu, R)$ calculated in BBKS. Appel & Jones (1990; hereafter AJ) attempted to formally derive the scale function of peaks with fixed density contrast. These authors assumed, for simplicity, that points which are peaks on a given scale keep on being peaks when the scale is changed, which is obviously not true in general. In changing the scale, the spatial location of peaks also changes. As a consequence, there is no obvious connection between peaks on different scales and, what is more important, between their respective density contrasts. In addition, the scale function derived in this way is not corrected for the important cloud-in-cloud effect affecting it. More recently, Bond & Myers (1993a, 1993b) have proposed a new method, the so-called "peak patch formalism", to obtain the mass function of objects. This follows the correct strategy for peaks, the cloud-in-cloud effect is corrected for, and a more accurate collapse dynamics than the spherical model is used. However, this new method is rather a numerical simulation; it does not provide us with any analytical or semianalytical expression for the mass function as wanted.

In the present paper we give a fully justified formal derivation of the theoretical mass function of objects relying just on the peak model ansatz. This derivation draws inspiration from (and, in fact, is very close to) AJ's. The main differences are 1) we do not assume that the spatial locations of peaks remain fixed when the scale is changed, but let them vary, and 2) we correct the resulting scale function for the nesting of collapsing clouds. These improvements with respect to AJ and the determination of unambiguous $M(R)$ and $\delta_c(t)$ relations are possible thanks to the development, in § 2, of a new formalism, hereafter referred to as the "confluent system formalism" (see also Salvador-Solé & Manrique 1994), which is able to follow the filtering evolution of peaks. In § 3 we apply this formalism to derive the scale function of peaks with fixed density contrast, and to correct it for the cloud-in-cloud effect. In § 4 we determine the form of the $M(R)$ and $\delta_c(t)$ relations which are consistent with the peak model for different power spectra of the density field in an Einstein-de Sitter universe. Finally, we derive, in § 5, the mass function of objects in these cosmogonies. Other important quantities connected with the detailed growth history of objects are calculated in paper II (Manrique & Salvador-Solé 1995).

## 2. THE CONFLUENT SYSTEM FORMALISM

In gravitational clustering one can make the practical distinction between accretion and merger. Accretion is, by definition, a continuous and differentiable process in time. For any accreting object of given mass $M$ another object of mass $M + dM$ can be found which subtends the former. In contrast, merger is a discontinuous event. There is a discrete gap $\Delta M$ in mass values in which no object subtending the matter of some given initial one can be found. This gap delimits the merging object whose mass evolution is being followed from the object resulting from the merger. According to the peak model ansatz enunciated above, peaks in the smoothed density field rearrange, with decreasing overdensity, essentially as objects do in time through accretion and mergers. Therefore, we can identify events analogous to accretion and merger in the filtering process in a very straightforward way by means of the correspondence between objects with increasing mass along the increasing time $t$ and those peaks tracing them at $t_i$ with decreasing density contrast $\delta$ when the filtering scale $R$ is increased.

## 2.1. Filtering Accretion

A point which is a peak on scale $R$ is not so, in general, on scale $R + \Delta R$, with $\Delta R$ positive and arbitrarily small. To guarantee that a peak on scale $R + \Delta R$ traces the same accreting object as the peak on the initial scale $R$ at the times corresponding to their respective density contrasts the separation between both points must be, at most, of the order of $\Delta R$. In this manner, the collapsing cloud associated with the peak on scale $R + \Delta R$ will include the volume (mass) subtended by the collapsing cloud associated with the peak on scale $R$. Furthermore, this proximity condition is not only necessary for the identification of peaks on contiguous scales, but also sufficient. Indeed, as readily seen from the Taylor series expansion of the density gradient around a density maximum, there cannot be more than one peak on scale $R + \Delta R$ in the neighborhood of any point which was a peak on scale $R$. (See § 2.2 for the case that no identification is possible for a given peak of scale $R$.)

This identification allows us to draw a $\delta$ vs. $R$ diagram similar to that obtained by Bond et al. (1991) in the excursion set formalism but for the fact that, in our diagram, each trajectory $\delta(R)$ is attached to one individual object or, what is the same, to *the changing peaks tracing it* in the filtering process instead of to one fixed point. To construct this diagram we must find all peaks on a given scale $R$ in some arbitrary volume, increase the scale by $\Delta R$, find the new peaks on the scale $R + \Delta R$, and identify each of them with one of the peaks on scale $R$, repeating the process from the largest scale reached at each step as many times as necessary. The continuous curves $\delta(R)$ determined by each series of identified peaks (disregarding their changing spatial location) represent the trajectories followed by peaks "evolving" through filtering accretion and trace the time evolution of the mass of bound virialized objects as they accrete matter.

The density contrast of an evolving peak on scale $R + \Delta R$ is, to first order in $\Delta R$, simply given by $\delta + \partial_R \delta \, \Delta R$ in terms of the values of the random variables $\delta$ and $\partial_R \delta$ at the same evolving peak on scale $R$. To see this one must simply take the Taylor series expansion of the density contrast of the former peak around the position $\mathbf{r}$ and scale $R$ of the latter and take into account that, according to our identification criterion, we have $O(|\Delta \mathbf{r}|^2) \leq O[(\Delta R)^2]$. Notice that there is, indeed, no first order term in $\Delta \mathbf{r}$ in that series expansion owing to the null density gradient in peaks. We therefore conclude that the total derivative $d\delta/dR$ of a peak trajectory in the $\delta$ vs. $R$ diagram coincides with the partial derivative $\partial_R \delta$ of the respective peak currently at $(R, \delta)$.

We are now ready to check the self-consistency of the peak model ansatz on which the previous natural identification criterion among peaks on different scales is based. As accreting objects evolve in time their mass obviously increases. But the mass is an increasing function of $R$, while the time a decreasing function of the density contrast. Consequently, peaks on scale $R + \Delta R$ must, for consistency, have smaller density contrast than those identified with them on scale $R$ or, equivalently, the total derivative $d\delta/dR$ of peak trajectories must be negative. Let us see whether this is really satisfied.

By writting the scale derivative of the density contrast smoothed with any spherical window $W(r^2/R^2)$ of scale $R$, $\delta(\mathbf{r}, R)$, in terms of the Fourier transform of the unfiltered field, $\delta(\mathbf{k}, 0)$, we have

$$\partial_R \delta(\mathbf{r}, R) = -\frac{R}{(2\pi)^3} \int_{-\infty}^{\infty} d^3k \, k^2 \, \delta(\mathbf{k}, 0) \, J(k^2 R^2) \exp(-i\mathbf{k}\mathbf{r}), \tag{5}$$

with $J(k^2 R^2)$ equal to $-2[\partial W(k^2 R^2)/\partial(k^2 R^2)]$ and $W(k^2 R^2)$ the Fourier transform of the smoothing window. Relation (5) can be rewritten in the form

$$\partial_R \delta(\mathbf{r}, R) = R \nabla^2 [\delta(\mathbf{r}, 0) * J(r^2/R^2)], \tag{6}$$

with $J(r^2/R^2)$ the inverse Fourier transform of $J(k^2 R^2)$ and $*$ denoting the convolution product. For a Gaussian window, i.e., $W(r^2/R^2) \equiv \exp[-r^2/(2R^2)]$, we have $J(k^2 R^2) = W(k^2 R^2)$ and equation (6) leads to the equality $\partial_R \delta(\mathbf{r}, R) = R \nabla^2 \delta(\mathbf{r}, R)$. This implies that $\partial_R \delta(\mathbf{r}, R)$ and, consequently, $d\delta/dR$ are automatically negative for peaks as needed. However, for any other window $J(k^2 R^2)$ is different from $W(k^2 R^2)$ and the sign of $\nabla^2 [\delta(\mathbf{r}, 0) * J(r^2/R^2)]$ is not determined by that of $\nabla^2 [\delta(\mathbf{r}, R)]$, but depends on the particular density distribution around each point $\mathbf{r}$. Thus, condition $d\delta/dR = \partial_R \delta(\mathbf{r}, R) < 0$ is not guaranteed for every peak. We are therefore led to the conclusion that the shape of the window used is crucial for the self-consistency of the peak model ansatz: only the Gaussian window is able to recover such a fundamental property of gravitational clustering as the systematic growth, by accretion, of the mass of objects in any realistic density field. This might explain the good behavior of the Gaussian window in $N$-body simulations of structure formation from peaks (Mellot, Pellman, & Shandarin 1993; see also Katz, Quinn, & Gelb 1993). It is interesting to note that the characteristics of the density distribution in the real universe causing the departure from spherical collapse also make the peak model ansatz inconsistent with the filtering by means of a top hat window. If such a density field were spherically symmetric and outwards decreasing around **r**, then the condition $\partial_R \delta(\mathbf{r}, R) < 0$ would be clearly fulfilled by the top hat window (and many others depending on the particular density profile). However, in the real universe, the Gaussian window is the only one that guarantees this condition. Herafter, we adopt this particular filter.

## 2.2. Filtering Mergers

As a peak evolves through filtering accretion the volume of its associated collapsing cloud increases. Provided there is full coverage of space by collapsing clouds associated with peaks of fixed overdensity, that volume increase makes peaks progressively become located inside the collapsing clouds of others with identical density contrast but larger filtering scale. This is but the well-known cloud-in-cloud effect which must be corrected for if we want that the scale function of peaks at a fixed overdensity reflects the mass function of virialized objects at the corresponding time. Notice that, the non-nested peaks that remain at each fixed overdensity will provide the exact coverage of space since just those causing the excessive coverage of space at any $\delta$ have been removed. Hereafter, we focus on the filtering evolution of non-nested peaks, as efficient tracers of virialized objects.

The nesting-corrected $\delta$ vs. $R$ diagram contains a set of continuous peak trajectories suddenly truncated when their respective evolving peaks become located inside the collapsing cloud associated with any other peak with identical density contrast but a larger filtering scale. Since the volume (mass) of the collapsing cloud associated with an accreting peak which becomes nested is covered by that of the host peak, we can think about the former as evolving into the latter (becoming part of it) through a discrete horizontal jump in the nesting-corrected $\delta$ vs. $R$ diagram. Such discrete jumps among peak trajectories towards larger scales along the line of fixed density contrast reflect discrete mass increases, at a fixed time, of the virialized objects they trace. Therefore, the nesting of, until then, non-nested peak trajectories can be naturally identified with mergers. Notice that the key assumption in this identification has been that collapsing clouds associated with non-nested peaks of given fixed density contrast yield the exact coverage of space. This is ensured by the peak model ansatz itself: an incomplete (or excessive) coverage of space by collapsing clouds associated with non-nested peaks is not allowed because this would overestimate (underestimate) the density of objects at the corresponding $t$ inferred from counting their respective seeds. Thus, our identification of filtering mergers also directly follows from the peak model ansatz.

But, apart from becoming nested, peak trajectories in the nesting-corrected $\delta$ vs. $R$ diagram can also disappear or appear. This is due to the fact that, as pointed out in § 2.1, the identification among peaks on different scales is not always possible. An infinitesimal increase in $R$ can make a continuous peak trajectory disappear (there is no peak on scale $R + \Delta R$ in the close neighborhood of the peak on scale $R$) or a new continuous peak trajectory appear (there is no peak on scale $R$ in the close neighborhood of the peak on scale $R + \Delta R$). When a peak appears (without being nested; otherwise the event would go unnoticed in the nesting-corrected $\delta$ vs. $R$ diagram) some peaks with identical density contrast and smaller scale automatically become nested into it. These filtering events therefore trace the *formation* of new virialized objects from the merger of smaller ones. However, when a peak disappears before becoming nested (otherwise we could ignore the event) the volume (mass) associated with it will be necessarily covered by that of collapsing clouds associated with other peaks with identical density contrast and *smaller* scale. This will produce the split of the former peak into the latter ones. $N$-body simulations also find sporadic splits in the *gravitational evolution of peaks* (van de Weygaert & Babul 1993). However, gravitational splits of peaks take place prior to collapse, while filtering splits of peaks refer to virialized objects. Thus, this latter kind of filtering events has no natural counterpart in the gravitational evolution of virialized objects. Moreover, these are not the only unrealistic events we can find in the filtering evolution of peaks. Nested peaks can also leave their host clouds yielding, in this manner, a different kind of split of peak trajectories. Thus, in contrast with virialized objects which tend to progressively cluster with each other without exception, peaks tend to come together as we diminish the critical overdensity but they also sporadically split into pieces. This reflects the limitations of the peak model to provide an exact description of the growth history of individual objects. Yet, we are not concerned with the detailed evolution of *individual* objects, but rather in the *statistical* description of the clustering process they follow. In fact, for the confluent system formalism to provide an acceptable clustering model we only need that both the net amount of peaks becoming nested (after substraction of those leaving their host clouds) and the net amount of peaks appearing as the result of mergers (after substraction of those disappearing and breaking into small pieces) are positive at any location of the nesting-corrected $\delta$ vs. $R$ diagram. In the present paper, we focus on obtaining the scale function of non-nested peaks at a fixed density contrast. In paper II, we calculate those net amounts and show the statistical validity of the confluence system formalism, at least for density fields leading to hierarchical clustering (in the bottom-up fashion), i.e., those in which $\sigma_0$ is a decreasing function of $R$.

## 2.3. The Confluent System Diagram

The nesting-corrected $\delta$ vs. $R$ diagram of such a hierarchical clustering looks like the idealized one plotted in Figure 1. To avoid crowding we have drawn just a few trajectories to illustrate the general behavior of the diagram one would obtain from a fair sample volume of the universe. As can be seen, this diagram differs from the analogous one obtained in the excursion set formalism (Bond et al. 1991) in three main aspects: 1) all trajectories have the same monotonous trend of decreasing density contrast with increasing scale, 2) they all have a finite continuous extent limited by mergers, and 3) the number of trajectories decreases with increasing filtering scale. Mergers reduce, indeed, the total number of surviving peaks. At $R = 0$ we have a large, usually infinite, number of trajectories, while for $R$ tending to infinity we end up with just one trajectory approaching to $\delta = 0$. This is the reason why we call this diagram the confluent system of peak trajectories. From Figure 1 it is apparent that the variation in the density of peaks above $\delta_c$ between scales $R$ and $R + \Delta R$ is not equal to the density of peaks upcrossing the $\delta_c$ line in this range of scales (nor is the mass of the collapsing clouds associated with them). Large horizontal skips along the $R$ axis direction caused by mergers also contribute to this variation, which invalidates equation (3). Note also that, owing to these horizontal skips, there are peaks missing on every scale $R$ making the integral of the *average mass* of collapsing clouds associated with peaks *at a fixed scale* be different from the mean density of the universe.

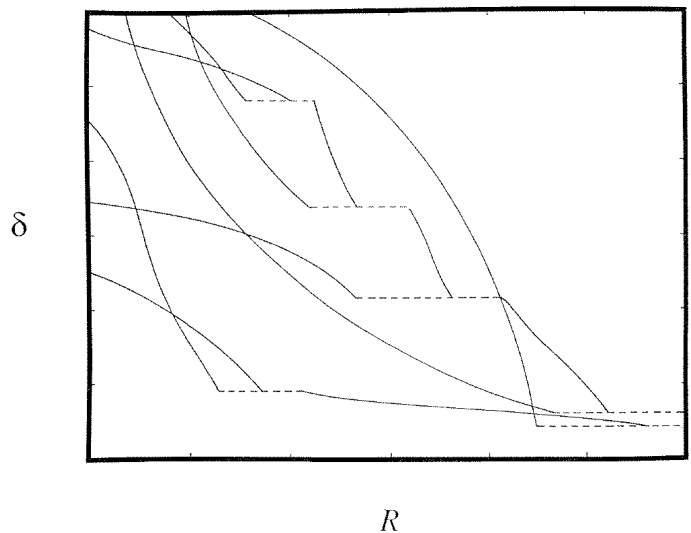

Fig. 1.— *Idealized confluent system of peak trajectories for a limited sample volume. In full line, the continuous filtering evolution of peaks tracing accretion by the corresponding virialized objects. In dashed lines, discontinuities tracing mergers of similarly massive objects.*

We want to stress that, as clearly stated at the begining of this section, the distinction between accretion and merger (in gravitational as well as in filtering evolution) is a practical one *which makes only sense from the point of view of any particular evolving object or peak*. In particular, what is a merger for one object can be either a merger or part of an accretion process for any other one partaking of the same event. This is shown in Figure 1. Peaks *resulting* from mergers can be either evolving through continuous accretion or forming depending on whether or not they can be identified with peaks at an infinitesimally larger $\delta$. This reflects a well-known fact in gravitational clustering: if some object partaking of a merger is massive enough relative to any other merger partner its relaxed state is not essentially altered by the event and one can keep on identifying it with the resulting object. Thus, from the point of view of such a massive object, the merger is a simple accretion. Conversely, from the point of view of the small accreted object which is destroyed in the event, the process is seen as a true merger of that object with a much more massive one. Of course, this latter phenomenon also has its counterpart in the filtering evolution of peaks: the exact coverage of space by collapsing clouds guarantees that the increase, with decreasing $\delta$, of the volume (mass) of clouds associated with any accreting peak is made at the expense of the volume (mass) of the collapsing clouds associated with those peaks which become nested into it. (This point of view of accretion is not systematically represented, however, in Fig. 1 because this would yield a continuous crowd of horizontal dashed lines starting from very small scales.) This shows that the analogy about mergers, accretion, and their interconnection, between gravitational and filtering evolutions is complete. It is important to note that the ambivalence of some processes of mass increase depending on the point of view of the particular object whose evolution is followed far from being a drawback of the confluent system formalism is at the base of the great potential of this tool. For example, in the excursion set formalism where there is no such ambivalence, accretion can only be treated as a series of mergers with very tiny objects (see, e.g., Lacey & Cole 1993). But, then, one cannot naturally distinguish such important events as the formation or the destruction of an object because there is no clear difference, from any point of view, between the mergers which characterize those notable events and all the mergers which constitute of the accretion process experienced by the object during its life. Thus, the possibility of following the mass increase of any given object by making the natural distinction between accretion and merger events is an important characteristic of the confluent system formalism with notable practical applications (see paper II).

## 3. THE SCALE FUNCTION

### 3.1. The Scale Function of Peaks with Fixed Density Contrast

– 13 –

To compute the density of peak trajectories upcrossing the $\delta_c$ line in infinitesimal ranges of scale we must calculate the density of peaks on scale $R$ with density contrast larger than $\delta_c$ which *evolve* into peaks with density contrast equal to or lower than that value on scale $R + \Delta R$, with $\Delta R$ positive and arbitrarily small. That is, we must compute the density of peaks on scale $R$ satisfying: $\delta_c < \delta$ and $\delta_c \geq \delta + d\delta/dR\, \Delta R$. From the results of § 2 we have that these two constraints can be expressed as

$$\delta_c < \delta \leq \delta_c - \nabla^2 \delta\, R\, \Delta R. \tag{7}$$

This coincides with the procedure followed by AJ. The difference between both approaches is that, contrary to AJ, *we do not assume that points which are peaks on scale $R$ keep on being peaks on scale $R + dR$*. We have just taken into account the identification criterion among peaks on different scales (with different locations, in general). This determines (see § 2) the density contrast of any peak at $R + \Delta R$ in terms of that of the peak at $R$ identified with it and, hence, the condition for any *evolving peak* to cross the threshold $\delta_c$.

Now, following BBKS step by step, we can readily calculate the mean density of peaks on scale $R$ satisfying the constraint given by equation (7) by taking the mean for the joint probability function $P(\delta, \boldsymbol{\eta} = 0, \zeta_A)$ of the full random density field of these peaks. Variables $\eta_i$ and $\zeta_A$ ($A = 1, 2, 3, 4, 5, 6$ stand for $ij = 11, 22, 33, 23, 13, 12$, respectively) are the first and second order cartesian derivatives of the mass density field smoothed on scale $R$, respectively. To calculate that mean it is convenient to use the new variables $\nu \equiv \delta/\sigma_0$, $x \equiv -(\zeta_1 + \zeta_2 + \zeta_3)/\sigma_2$, $y \equiv -(\zeta_1 - \zeta_3)/(2\sigma_2)$, and $z \equiv -(\zeta_1 - 2\zeta_2 + \zeta_3)/(2\sigma_2)$, with $\sigma_j$ the $j$th spectral momentum

$$\sigma_j^2(R) = \int_0^\infty \frac{dk\, k^{2(j+1)}}{2\pi^2}\, P(k) \exp(-k^2 R^2), \tag{8}$$

where $P(k)$ is the power spectrum of the density field. The only difference with respect to the procedure followed by BBKS is that we must include the extra factor $\delta(x - x_0)$ in the calculation of the mean density of peaks in order to obtain a function of the variables $\nu_0$ and $x_0$ ($x_0 > 0$). This leads to

$$\mathcal{N}_{pk}(\nu, x, R)\, d\nu\, dx = \frac{\exp(-\nu^2/2)}{(2\pi)^2 R_*^3 [2\pi(1-\gamma^2)]^{1/2}}\, f(x) \exp\left[-\frac{(x - \gamma\nu)^2}{2(1-\gamma^2)}\right] d\nu\, dx \tag{9}$$

(dropping subindexes 0), with $f(x)$ a function given by BBKS (eq. [A15]), $\gamma \equiv \sigma_1^2/(\sigma_0 \sigma_2)$, and $R_* \equiv \sqrt{3}\sigma_1/\sigma_2$. The density function of peaks satisfying condition (7) is therefore

$$N_{pk}(R, \delta_c) = \lim_{\Delta R \to 0} \frac{1}{\Delta R} \int_0^\infty dx \int_{\nu_c}^{\nu_c + [\sigma_2(R)/\sigma_0(R)]\, x\, R\, \Delta R} d\nu\, \mathcal{N}_{pk}(\nu, x, R), \tag{10}$$

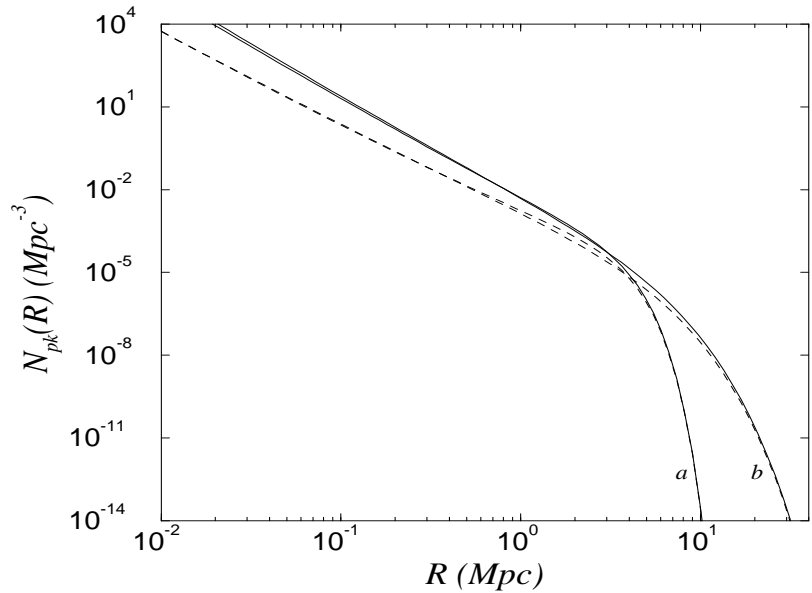

Fig. 2.— *Comparison between the scale function of peaks derived here (full line), and the one guessed by Bond (1989) (dashed line) corresponding to the present epoch in a univers with $\Omega = 1$, $\Lambda = 0$, $h = 0.5$ and a density field endowed with CDM (a) or $n = -2$ power-law (b) power spectrum, normalized to the present rms density fluctuation inside a sphere of $8h^{-1}$ Mpc equal to 0.67. The value of the critical density $\delta_{c0}$ used for each power spectrum is the same as in Fig. 4.*

with $\nu_c \equiv \delta_c/\sigma_0(R)$.

By integrating over $\nu$ and $x$ and dividing by $\Delta R$ we finally obtain at $\Delta R \to 0$

$$N_{pk}(R,\delta_c)\,dR = \frac{H(\gamma,x_*)}{(2\pi)^2\,R_*^3}\,\exp\left(-\frac{\nu_c^2}{2}\right)\frac{\sigma_2(R)}{\sigma_0(R)}\,R\,dR, \qquad (11)$$

with $x_* \equiv \gamma\nu_c$, and

$$H(\gamma,x_*) = \int_0^\infty x\,f(x)\,\frac{\exp\left[-\dfrac{(x-x_*)^2}{2\,(1-\gamma^2)}\right]}{[2\pi\,(1-\gamma^2)]^{1/2}}\,dx. \qquad (12)$$

The scale function (11) coincides with that given by AJ (except for a factor two in their final expression). It is also very similar to the expression given by equation (4) guessed by Bond (1989). Indeed, an alternate expression for equation (11) in terms of the density function $\mathcal{N}_{pk}(\nu_c, R)$ equal to the integral of $\mathcal{N}_{pk}(\nu_c, x, R)$ over the whole range of positive values of $x$ is

$$N_{pk}(R,\delta_c)\,dR = \mathcal{N}_{pk}(\nu_c, R)\,\frac{\sigma_2(R)}{\sigma_0(R)}\,<x>\,R\,dR, \qquad (13)$$

with

$$<x> \equiv \frac{\int_0^\infty x\,f(x)\exp\left[-\frac{(x-x_*)^2}{2(1-\gamma^2)}\right]dx}{\int_0^\infty f(x)\exp\left[-\frac{(x-x_*)^2}{2(1-\gamma^2)}\right]dx}. \qquad (14)$$

While, the relation

$$\frac{1}{\sigma_0}\frac{d\sigma_0}{dR} = -R\frac{\sigma_1^2}{\sigma_0^2} \qquad (15)$$

valid for the Gaussian filtering allows us to write equation (4) in the form

$$N_{pk}(R,\delta_c)\,dR = \mathcal{N}_{pk}(\nu_c, R)\frac{\sigma_2(R)}{\sigma_0(R)}x_*\,R\,dR. \qquad (16)$$

Equations (13) and (16) only differ by the effective scaled Laplacian $x$ appearing in both expressions. This difference introduces, however, a notable deviation between the two functions at the small-scale end (see Fig. 2). For power-law power spectra, $P(k) \propto k^n$ ($-3 < n < 4$), $\gamma$ is constant, $R_*$ is proportional to $R$, and the function $H$ is constant for small $R$. Since $\sigma_2/\sigma_0$ is proportional to $R^{-2}$ the logarithmic slope at small $R$ of the scale function (11) or (13) is therefore equal to $-4$. As pointed out by AJ, such a steep slope makes the mass integral diverge for collapsing clouds with $M$ proportional to $R^3$. (This is the same to say that there is a divergent coverage of space by such collapsing clouds). But we do not know the actual relation $M(R)$. Moreover, before applying the scale function (11) or (13) to objects we must correct it for the cloud-in-cloud effect.

### 3.2. Correction for the Cloud-in-Cloud Effect

To perform this correction we must first compute the density of peaks at fixed density contrast, subject to the condition of being located in some particular background. Following just the same reasoning as in § 3.1 but from the conditional density $\mathcal{N}_{pk}(x,\nu,R|\nu_b,R_b)\,dx\,d\nu$ given by BBKS instead of the density $\mathcal{N}_{pk}(x,\nu,R)\,dx\,d\nu$ (eq. [9]), one is led to the following density of peaks with density contrast $\delta_c$ on scales between $R$ and $R+dR$ at points with density contrast $\delta_b$ on scales $R_b$

$$N_{pk}(R,\delta_c|R_b,\delta_b)\,dR = \frac{H(\tilde{\gamma},\tilde{x}_*)}{(2\pi)^2 R_*^3}\frac{\exp\left[-\frac{(\nu_c-\epsilon\nu_b)^2}{2(1-\epsilon^2)}\right]}{(1-\epsilon^2)^{1/2}}\frac{\sigma_2(R)}{\sigma_0(R)}R\,dR, \qquad (17)$$

where $\tilde{x}_*$ is defined as $\tilde{\gamma}\tilde{\nu}$, with

$$\tilde{\gamma}^2 = \gamma^2\left[1+\epsilon^2\frac{(1-r_1)^2}{1-\epsilon^2}\right], \qquad \tilde{\nu} = \frac{\gamma}{\tilde{\gamma}}\left(\frac{1-r_1}{1-\epsilon^2}\right)\left[\nu_c\left(\frac{1-\epsilon^2 r_1}{1-r_1}\right)-\epsilon\nu_b\right], \qquad (18)$$

in terms of the spectral parameters $\epsilon \equiv \sigma_{0h}^2/(\sigma_0 \sigma_{0b})$, and $r_1 \equiv \sigma_{1h}^2 \sigma_0^2/(\sigma_{0h}^2 \sigma_1^2)$, with $\sigma_{jh}(R, R_b)$ defined, for a Gaussian window, just as $\sigma_j$ (eq. [6]) but for the rms average scale $R_h \equiv [(R^2 + R_b^2)/2]^{1/2}$. The conditional density given in equation (17) can also be expressed as

$$N_{pk}(R, \delta_c | R_b, \delta_b) \, dR = \mathcal{N}_{pk}(\nu_c, R | \nu_b, R_b) \, \sigma_2(R) <\widetilde{x}> R \, dR, \tag{19}$$

in terms of the conditional density function $\mathcal{N}_{pk}(\nu, R | \nu_b, R_b)$ calculated in BBKS and the function $<\widetilde{x}>$ given by equation (14) now in terms of $\tilde{\gamma}$ and $\tilde{x}_*$ instead of $\gamma$ and $x_*$.

We are now ready to obtain the master equation yielding the scale function of peaks at a fixed density contrast $\delta_c$ corrected for the cloud-in-cloud effect. The probability that a point is located inside the collapsing cloud associated with a non-nested peak of density contrast $\delta_c$ on some scale in the range between $R_b$ and $R_b + dR_b$ is $\rho^{-1} M(R_b) N(R_b, \delta_c) \, dR_b$, where $\rho^{-1} M(R_b)$ is the volume of the collapsing cloud associated with the non-nested peak, and $N(R_b, \delta_c) \, dR_b$ is the unknown scale function of non-nested peaks. Given the meaning of the conditional density (19) we are led to the following relation

$$N(R, \delta_c) = N_{pk}(R, \delta_c) - \frac{1}{\rho} \int_R^\infty dR_b \, M(R_b) \, N(R_b, \delta_c) \, N_{pk}(R, \delta_c | R_b, \delta_c). \tag{20}$$

Equation (20) is a Volterra type integral equation of the second kind for the nesting-corrected scale function of peaks $N(R, \delta_c)$. According to the theory of integral equations, there exists a unique solution which can be obtained, numerically, by iteration from the initial approximate solution $N_{pk}(R, \delta_c)$ given by equation (11). However, to solve equation (20) we must previously determine the function $M(R)$ in the kernel giving the mass of collapsing clouds associated with peaks at scales between $R$ and $R + dR$. In the spherical collapse model $M(R)$ is equal to $4\pi/3 \, \rho \, R^3$, that is, the mass subtended by a top hat window of scale $R$. But we know that this simple model does not apply. In a similar manner as the Gaussian window is better suited than the top hat one (§ 2), the right function $M(R)$ to use can notably deviate from that expression. On the other hand, to obtain the mass function $N(M, t) \, dM$ from the scale function $N(R, \delta_c) \, dR$ solution of equation (20), we also need the relation $\delta_c(t)$ which, for identical reasons, can notably deviate from the usual expression for the spherical collapse.

## 4. DYNAMICAL CONSTRAINTS ON THE $M(R)$ AND $\delta_c(t)$ RELATIONS

### 4.1. The Mass vs. Scale Relation

In what follows, we will only consider the case of an Einstein-de Sitter universe ($\Omega = 1$, $\Lambda = 0$). If the density field is endowed with a power-law power spectrum there is no privileged time nor scale. So the scale function of peaks at a fixed density contrast must be self-similar. This means that if we define, at any epoch, a characteristic length $R_c$ from some arbitrarily fixed value of any physically distinguishable (although not privileged) quantity such as the amplitude of density fluctuations, through, say, $\sigma_0(R_c) = \delta_c$, then any quantity reporting to that characteristic length must be invariant in time. One of such quantities is the number of peaks inside the volume $R_c^3$ with density contrast $\sigma_0(R_c)$ (equal to $\delta_c$) on scales $R/R_c$ in an infinitesimal range. The density function of peaks with $\delta_c$ on scales in units of $R_c$ is equal to $R_c$ times the function $N_{pk}(R, \delta_c)$ given by equation (11). Thus, by multiplying this density by $R_c^3$ and writting all spectral moments involved in terms of $R$, $R_c$, and $n$ we are led to

$$\mu_{pk}(R, \delta_c) = \frac{(n+5)^2(n+3)^{1/2}}{12\sqrt{6}\,(2\pi)^2}\, H\left[\frac{n+3}{n+5}\left(\frac{R}{R_c}\right)^{\frac{n+3}{2}}\right]\left(\frac{R}{R_c}\right)^{-4}\exp\left[-\frac{1}{2}\left(\frac{R}{R_c}\right)^{n+3}\right]. \qquad (21)$$

As can be seen, the right-hand side of equation (21) depends on $R$ and $\delta_c$ just through the ratio $R/R_c$. Hence, $\mu_{pk}$ is time-invariant. Likewise, the number of *non-nested* peaks, $\mu(R, \delta_c)\,d(R/R_c)$, inside the same comoving volume with density contrast $\sigma_0(R_c)$ (or $\delta_c$) on scales $R/R_c$ in an infinitesimal range must also be invariant. From equation (20) we have

$$\mu(R, \delta_c) = \mu_{pk}(R/R_c) - \frac{1}{\rho}\int_{R/R_c}^{\infty} d(R_b/R_c)\,M(R_b)\,\mu(R_b/R_c)\,R_c\,N_{pk}(R, \delta_c|R_b, \delta_c). \qquad (22)$$

By writting all spectral moments involved in the integrant on the right-hand side of equation (22) in terms of $R$, $R_b$, $R_c$, and $n$, we obtain a function of $R/R_c$ and $R_b/R_c$ times an extra factor $R^{-3}$ or, equivalently, an extra factor $R_c^3$. Thus, $\mu(R, \delta_c)$ in equation (22) will be invariant as required *provided only that the mass $M(R_b)$ is proportional to $R_b^3$*. Indeed, by multiplying and dividing this integrant by $R_c^3$ we obtain a function of just $R/R_c$ and $R_b/R_c$ and, by integrating it, a function of $R/R_c$ as $\mu_{pk}$. We therefore arrive to the conclusion that the dynamical consistency of the scale function solution of equation (20) implies

$$M(R) = \rho\,(2\pi)^{3/2}\,[q\,R]^3, \qquad (23)$$

with $q$ an arbitrary constant likely dependent on the spectral index $n$. The natural volume subtended by a Gaussian window of scale $R$ is $(2\pi)^{3/2}\,R^3$. So the meaning of $q$ in equation (23) is simply the ratio between the true Gaussian length of the collapsing cloud associated with a peak and the filtering scale used to find it. (In principle, the invariance of $\mu(R, \delta_c)$ would also hold if $q$ were a function of $R/R_c$. But this more general dependence of $q$ is not allowed because $M$ can only depend on $R$, not on $\delta_c$.) A third invariant function which is then also guaranteed is the mass fraction in objects with masses, in units of $M_c \equiv M(R_c)$, in an infinitesimal range. This is, by the way, the only invariant function one has in the usual PS approach. But from equation (1) we see that the function $M(R)$ cancels out in the explicit expression of this invariant. This is the reason why one cannot use a similar reasoning as above to constrain the form of $M(R)$ in that approach.

Equation (23) is only valid in the scale-free case. Nonetheless, every power spectrum can be approximated by a set of different power-laws in specific finite ranges of the scale. Thus, under the reasonable assumption that the dynamics of the collapse for a fluctuation of any given scale only depends (statistically) on the distribution of density fluctuations on similar scales, the form of $M(R)$ for non-power spectra will approximately follow equation (23). Some departure from that simple law cannot be avoided, for example, in the case of the CDM spectrum if the values of the constant $q$ for the two power laws giving the asymptotic regimes at large and small scales are distinct. However, provided that the values of $q$ for different spectral indexes $n$ are not too different from each other, equation (23) will be a good approximation for some effective fixed value of $q$ dependent, in general, on the particular non-power-law spectrum used. Let us adopt, hereafter, this simplifying assumption and check its validity a posteriori.

There is still another constraint on the function $M(R)$. The nesting-corrected scale function must satisfy the normalization condition

$$1 = \int_0^\infty \frac{M(R)}{\rho} N(R, \delta_c)\, dR. \tag{24}$$

Equation (24) expresses the fact that the collapsing clouds associated with non-nested peaks yield the exact coverage of space. (Remember that this is equivalent to ask for any particle in the universe to be, at the time $t$, inside some virialized object with the appropriate mass.) Condition (24) should be automatically satisfied for any power-law spectrum. Indeed, as discussed at the end of § 3, collapsing clouds associated with peaks prior to the correction for the nesting effect yield, in this case, a divergent coverage of space. Hence, after correction for the nesting effect one should end up with the exact coverage. As a consequence, we expect the scale function $N(R, \delta_c)$ to be correctly normalized for whatever value of $q$. However, for non-power-law spectra, the coverage of space by collapsing clouds associated with uncorrected peaks may not diverge (for example, in the case of the CDM spectrum). Then, for any given value of $\delta_c$ there should be a unique effective value of $q$ yielding the correct normalization. But $M$ cannot depend on $\delta_c$. We are therefore led to the following conclusion: if the (effective) value of $q$ yielding the correct normalization is the same for every $\delta_c$ there will be a unique physically consistent solution of equation (20), while not, there will be no acceptable solution. It is worthwhile noting that, since the mass function given by equation (1) is always correctly normalized, the form of $M(R)$ in the PS approach, is not constrained by condition (24), either.

In summary, the existence of some consistent scale function $N(R, \delta_c)$ is only guaranteed in the scale-free case. Constant $q$ is then a free parameter. For non-power-law spectra it is hard to tell a priori whether or not there is some consistent solution. But if there is, the effective value of parameter $q$ is automatically fixed.

## 4.2. Overdensity vs. Collapse Time Relation

The previous arguments only concerned the scale function of peaks with fixed density contrast $\delta_c$ at $t_i$. Let us now turn to the corresponding mass function of objects at $t$. For this mass function to be well-defined, that is, independent of the arbitrary initial time $t_i$ chosen, $\delta_c$ must be proportional to $a(t_i)$. Indeed, on changing $t_i$ the values of the spectral parameters $\gamma$, $\sigma_1/\sigma_2$, and $\sigma_2/\sigma_0$ do not vary, the only variable affected being $\nu_c$ or, more exactly, $\sigma_0$ which appears dividing $\delta_c$ in the scale function (11) as well as in the conditional density of peaks in the kernel of equation (20). Since $\sigma_0$ changes as $a(t_i)$ $\delta_c$ must also vary as $a(t_i)$ in order to balance that change. Therefore, the most general form for the critical overdensity vs. collapse time relation is

$$\delta_c(t) = \delta_{c0}(t) \frac{a(t_i)}{a(t)}. \qquad (25)$$

For the scale-free case, any change in $t_i$ and $t$ determining the same cosmic expansion factor $a(t)/a(t_i)$ should go unnoticed. (There is no absolute reference to assess the shift produced in the normalization of the power spectrum, that is, in the value of $\sigma_0$.) Therefore, $\delta_{c0}$ in equation (25) must be constant and equation (25) reduces to the usual expression for the spherical collapse model except for the fact that the value of $\delta_{c0}$ can be different from 1.69 and possibly depend on the power index $n$. For non-power-law spectra, there is no obvious constraint on $\delta_{c0}(t)$. The spherical collapse model suggests that it should not be far from constant, but some slight departure is probable, just as for $q$ in the function $M(R)$ discussed above. However, for identical reasons, we will assume that $\delta_{c0}$ is approximately equal to some effective constant value dependent, in general, on the particular non-power-law spectrum used, and check a posteriori the validity of this approximation.

A last comment is in order concerning the constants $\delta_{c0}$ and $q$. In the PS approach and a power-law spectrum, there is a degeneracy in their values. One can take any of them fixed according to the spherical collapse model and adjust the value of the other one by, for instance, fitting the mass function of virialized objects obtained from $N$-body simulations. This is due to the fact that the mass fraction in virialized objects, the only time-invariant function we have in this case, depends on $q$ and $\delta_{c0}$ only through the characteristic mass $M_c = M(R_c)$ (see, e.g., Lacey & Cole 1994). As a consequence, every combination of the two parameters leading to the same value of $M_c$ yields the same mass fraction or, equivalently, the same family of mass functions for different times. In contrast, there is no such degeneracy in the peak model framework. The mass fraction in objects depends on $q$ and $\delta_{c0}$ not only through $M_c$, but also through $q$ separately. Indeed, the mass $M(R_b)$ appearing in the invariant number of non-nested peaks inside the volume $R_c^3$ (eq. [22]) depends only on $q$ and the situation does not improve when that number is multiplied by $M(R)/(\rho R_c^3)$ in order to obtain the mass fraction in objects. The fact that, in the peak model, a change in $q$ cannot be balanced by any change in $\delta_{c0}$ is not surprising since the parameter $q$ controls by itself the importance of the nesting effect. Indeed, the value of $q$ determines the invariant fraction of non-nested peaks, equal to the ratio of equations (22) and (21).

## 5. THE MASS FUNCTION OF OBJECTS

The resulting mass function of objects in an Einstein-de Sitter universe is

$$N(M,t)\,dM = N(R,\delta_c)\frac{dR}{dM}\,dM, \qquad (26)$$

with $N(R,\delta_c)$ given by equation (20), $M(R)$ by equation (23) for some unknown constant $q$, and $\delta_c(t)$ by equation (25) for $\delta_{c0}$ another unknown constant. Of course, the existence of any consistent solution is not yet guaranteed (see the end of § 4.1). Moreover, for any solution to be dynamically acceptable it must be close, for any value of $t$, to the PS mass function with top hat window and $\delta_{c0} = 1.69$ because this gives good fits to the mass function inferred from $N$-body simulations. The situation is specially critical in the case of non-power-law spectra because the similarity between both mass functions for different values of $t$ is not trivial (they are not self-similar), and there is just the free parameter $\delta_{c0}$ to be adjusted.

For a given power spectrum and some arbitrary fixed values of $t$ and $t_i$, we have solved equation (20) as indicated at the end of § 3 by choosing the value of $q$ that satisfies, for each different value of $\delta_{c0}$ tried, the normalization condition (24). In the case of power-law spectra we find that the normalization condition (24) is satisfied by any value of $q$ and $\delta_{c0}$, as expected. (Varying $\delta_{c0}$ is equivalent, for a fixed value of $t_i$, to vary $\delta_c$.) In the case of the CDM spectrum, the correct normalization is only obtained for one specific value of $q$ for each $\delta_{c0}$ tried, also as expected. What is most remarkable is that this value of $q$ is quite insensitive to $\delta_{c0}$ just as needed for our simplifying assumption on the shape of $M(R)$ in that case to be acceptable. As shown in Figure 3, $q$ exhibits only a 10 % variation around the value 1.45 for values of $\delta_{c0}$ expanding along two decades.

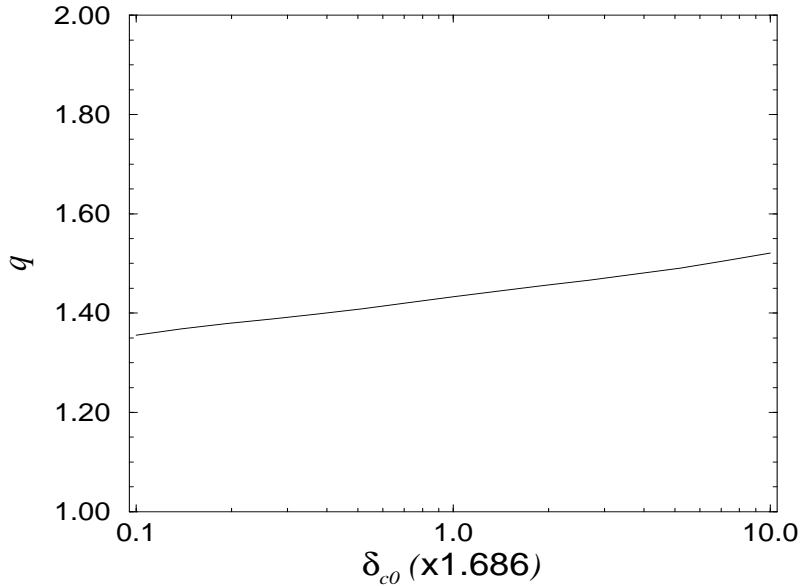

Fig. 3.— *Dependence on $\delta_{c0}$ of the parameter $q$ for the CDM spectrum assuming the same mass vs. scale relation as found for power-law espectra.*

It is worthwhile mentioning that the normalization condition (24) can be checked very accurately. Apart from numerical roundoff errors, there is a some uncertainty arising from the fact that to estimate the total mass integral we must extrapolate to $R = 0$ the scale function obtained down to some non-vanishing scale. In the case of power-law spectra, we cannot directly reach $R = 0$ because of the divergence there of the spectral moments. Yet, a log-log extrapolation gives an excellent approximation to the total mass integral which turns out to be correctly normalized up to an accuracy of $10^{-5}$. In the case of the CDM spectrum, one cannot trust the scale function at small scales because of the poorly known shape of the power spectrum at very large $k$ (all available analytic approximations are only valid up to some finite wavenumber). Nonetheless, the total mass integral is essentially controlled, in this case, by the well-determined large-scale end of the scale function so that the result is also reliable. After trying with different analytic approximations for the CDM power spectrum, all with the expected $k^{-3}$ asymptotical behavior, we can ascertain that the values of $q$ satisfying the normalization condition for any given $\delta_{c0}$ are correct within 1 % of accuracy.

In Figure 4 we plot, for the same power spectra as in Figure 2, the solutions obtained for the values of the parameters $q$ (if free) and $\delta_{c0}$ which give the best fit to the corresponding PS mass function. As can be seen, the similarity between both solutions in the CDM case for $\delta_{c0} \approx 6.4$ (and $q \approx 1.45$) is remarkable. This confirms that the relation $M(R)$ can be approximated, indeed, by equation (22). The similarity between both solutions also holds for any power-law spectrum tried. The solution plotted in Figure 4, corresponding to $n = -2$, is obtained for $\delta_{c0} \approx 8.4$ and $q \approx 1.45$. (Spectral indexes larger than $-2$ lead to values of $q$ slightly smaller than 1.45. So the equality in the preceding two values of $q$ is a mere coincidence.) Finally, Figure 4 also shows that these similarities hold for any value of $t$ as required. Thus, the validity of our simplifying assumption that $\delta_{c0}(t)$ in equation (24) is approximately constant in the CDM case is also confirmed.

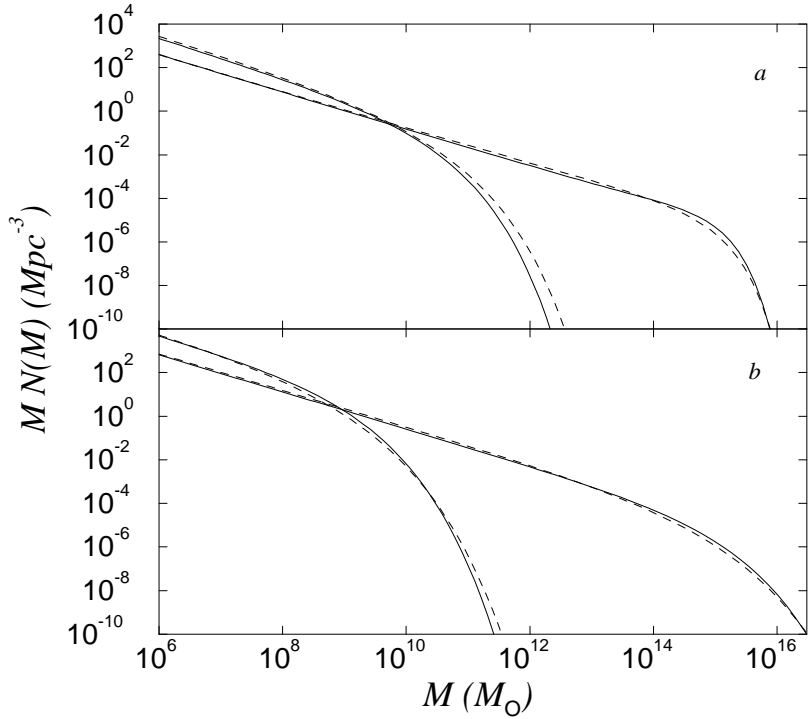

Fig. 4.— The final mass functions specified in §5 (see text for the values of $q$ and $\delta_{c0}$) for the same spectra as in Fig. 2 (full lines) compared with PS original, correctly normalized, mass functions for top-hat filtering and $\delta_{c0} = 1.69$ (dashed lines). In each panel we plot the solutions corresponding to two different epochs: the present time (curves reaching higher masses) and the time at which the cosmic scale factor was a tenth of its current value.

More accurate values for the constants $q$ and $\delta_{c0}$ require the direct fitting of the mass function inferred from $N$-body simulations. In any event, there is little doubt on the marked departure of $\delta_{c0}$ from 1.69. This is not only caused by the departure from the spherical collapse, but mainly by the values of $q$ found. These are large compared to the value of 0.64 yielding the same mass in objects for the Gaussian window as for the top hat one with identical scale $R$. But the collapse at the correct time of a large cloud requires (at least in the spherical collapse framework) a large value of the density contrast averaged just in the small central region. Hence, it is not surprising that large values of $\delta_{c0}$ are coupled to

– 23 –

large values of $q$, while the values of both parameters cannot be reduced because there is no degeneracy in the confluence system formalism.

To conclude we want to stress that the present derivation of the mass function of objects relying just on the validity of the peak model ansatz has turned out to be an important test for this latter model. The fact that it has been possible to derive a fully consistent, well justified, and dynamically acceptable mass function gives strong support to the statistical validity of that simple model of structure formation provided, of course, the use of the appropriate window and $M(R)$ and $\delta_c(t)$ relations.

This work has been supported by the Dirección General de Investigación Científica y Técnica under contract PB93-0821-C02-01. We thank J.M. Solanes for kindly reviewing the manuscript.